\documentclass[twocolumn,showpacs,preprintnumbers,amsmath,amssymb]{revtex4}
\usepackage{graphicx}% Include figure files
\usepackage{dcolumn}% Align table columns on decimal point
\usepackage{bm}% bold math

\begin{document}

%\preprint{}

\title {Second post-Newtonian Lagrangian dynamics of spinning compact binaries} % Force line breaks with \\

\author{Li Huang and Xin Wu}
\email{xwu@ncu.edu.cn} \affiliation{Department of
Physics $\&$ Institute of Astronomy, Nanchang University, Nanchang 330031, China}

%%\date{\today} % It is always \today, today,

\begin{abstract}

The leading-order spin-orbit coupling is included in a
post-Newtonian Lagrangian formulation of  spinning compact binaries,
which consists of the Newtonian term,
first post-Newtonian (1PN) and 2PN non-spin terms and 2PN spin-spin coupling.
This makes a 3PN spin-spin coupling occur in the derived Hamiltonian.
The spin-spin couplings are mainly responsible for chaos in the
Hamiltonians. However,
the 3PN spin-spin Hamiltonian is small and has different signs,
compared with the 2PN spin-spin Hamiltonian equivalent to the
2PN spin-spin Lagrangian.
As a result, the probability of the occurrence of chaos
in the Lagrangian formulation without the spin-orbit coupling
is larger than that in the Lagrangian formulation with the spin-orbit
coupling. Numerical evidences support the claim.

\end{abstract}

\pacs{04.25.Nx, 04.25.-g, 05.45.-a, 45.20.Jj}% PACS, the Physics and Astronomy
% Nonlinear dynamics and chaos 05.45.-a,
% 45.20.Jj Lagrangian and Hamiltonian mechanics  ,
%95.10.Fh Chaotic dynamics

\maketitle

\section{Introduction}
On February 11, 2016, the LIGO Scientific Collaboration and Virgo Collaboration
announced the detection of gravitational wave signals (GW150914),
sent out from the inspiral and merger of a pair of black holes with masses 36$M_{\odot}$ and 29$M_{\odot}$ [1].
The gravitational wave discovery directly confirmed a major prediction of Albert Einstein's 1915 general theory of relativity.
The LIGO project was originally proposed in the 1980s and its initial funding was approved in 1992. Since then,
the dynamics of spinning compact binaries has received more attention. A precise theoretical waveform template is necessary to match with
gravitational wave data. As a kind of description of the waveforms and dynamical evolution equations, the post-Newtonian (PN) approximation
to general relativity was used. Up to now, the PN expansion of the relativistic spinning two-body problem has provided
the dynamical non-spin evolution equations and the spin evolution equations to fourth post-Newtonian (4PN) order [2-7].

A key feature of the gravitational waveforms from a chaotic system is the extreme sensitivity to initial conditions [8-10],
and therefore the chaos is a possible obstacle to the method of
matched filtering. For this reason, several authors were interested in  the  presence or absence of
chaotic behavior in the orbital dynamics of spinning black hole pairs [11-17]. There were three debates about
this topic.

Sixteen years ago fractal methods were used to show that
the 2PN  harmonic-coordinates Lagrangian formulation of two spinning black holes admits chaotic behavior [11].
Here the Lagrangian  includes contributions from the Newtonian, 1PN and 2PN non-spin terms
and the effects of spin-orbit and spin-spin couplings.\footnote{The leading-order spin-orbit coupling is at 1.5PN order and
the leading-order spin-spin coupling is at 2PN order.} However, the authors of [12] made an opposite claim on ruling
out chaos in compact binary systems by finding no positive Lyapunov
exponents along the fractal of [11]. The authors of [18, 19] refuted this claim by finding positive Lyapunov exponents,
and pointed out that the reason for the false  Lyapunov
exponents obtained in Ref. [12] lies in using the Cartesian distance between
two nearby trajectories by continually rescaling the shadow trajectory. This is a debate
with respect to Lyapunov
exponents resulting in two different claims on the chaotic behavior
of comparable mass spinning binaries. In fact, the true reason for the discrepancy was found in Ref. [20]
and should be that different treatments of the Lyapunov exponents
were given in the three papers [12, 18, 19]. The authors of [12] used the stabilizing limit values as the
values of Lyapunov exponents, whereas those of [18, 19] used the slopes of the fit lines. It is clear that obtaining the limit values requires more
CPU times than obtaining the slopes.

Second debate is different descriptions of chaotic regions and parameter spaces.
It was reported in Ref. [21] that increasing the spin
magnitudes and misalignments leads to the transition to chaos, and the strength of chaos is the largest for the spins perpendicular to the orbital
angular momentum. However, an entirely different description from Ref. [22] is that chaos occurs mainly when
the initial spins are nearly antialigned with the orbital angular momentum for the
binary configuration of masses 10$M_{\odot}$ and 10$M_{\odot}$.
These  descriptions seem to be apparently conflicting, but
they can all be correct, as mentioned in Ref. [23]. This is because a complicated combination of all parameters and initial conditions
rather than single physical parameter or initial condition is responsible for yielding
chaos. No universal rule can be given to dependence of chaos on each parameter or initial condition.

Third debate relates to different dynamical behaviors of PN Lagrangian and Hamiltonian conservative systems at the same order.
In other words, the debate corresponds to a question whether the two formulations are equivalent.
The equivalence of Arnowitt-Deser-Misner (ADM) Hamiltonian and harmonic coordinate Lagrangian approaches
at 3PN order were proved by two groups [24, 25]. This result is still correct for approximation
accuracy to next-to-next-to-leading (4PN) order spin1-spin2 couplings [4].
Recently, the authors of [26] resurveyed this question. In their opinion,
this equivalence means that all the known results of the ADM Hamiltonian approach can be transferred to
the harmonic-coordinates Lagrangian one, or those of the harmonic-coordinates Lagrangian approach can be transferred to
the ADM Hamiltonian one. In fact, the two approaches are not exactly equal but are approximately related.
In this case, dynamical differences between the two approaches are not avoided.
For a two-black hole system with two bodies spinning,
the PN Lagrangian formulation of the Newtonian term and the 1.5PN spin-orbit coupling is non-integrable and possibly chaotic,
whereas the PN Hamiltonian formulation of the Newtonian term and the 1.5PN spin-orbit coupling is integrable and non-chaotic [26].
This is due to the difference between the two formulations, 3PN spin-spin coupling. That is,
the Lagrangian is exactly equivalent to the Hamiltonian plus the 3PN spin-spin term.
As a result, the 3PN spin-spin effect leads to the non-integrability of the Hamiltonian equivalent to the Lagrangian.
Ten years ago, similar claims were given to the two PN formulations of the two-black hole system with one bodies spinning [21, 27-29].
An acute debate has arisen as to whether the compact
binaries of one bodies spinning exhibit chaotic behaviour. A key to this question in Ref. [30] is as follows.
The construction of canonical, conjugate spin variables in Ref. [31] showed directly that
four integrals of the total energy and total angular momentum in an eight-dimensional phase
space of a conservative Hamiltonian equivalent to the
Lagrangian determine the integrability of the
Lagrangian. Precisely speaking, no chaos occurs in the
PN conservative Lagrangian and Hamiltonian formulations of comparable mass compact binaries
with one body spinning.

One of the main results of [26, 32] is that a PN Lagrangian approach at a certain order always exists a formal equivalent PN Hamiltonian.
This is helpful for us to study the Lagrangian dynamics using the Hamiltonian dynamics.
Following this direction, we shall revisit the 2PN ADM Lagrangian dynamics of two spinning black holes,
in which the Newtonian, 1PN and 2PN non-spin terms
and the 1.5PN spin-orbit and 2PN spin-spin contributions are included.
A comparison between the Lagrangian and related Hamiltonian dynamics will be made,
and a question of how the orbit-spin coupling exerts an influence on chaos
resulted from the spin-spin coupling in the Lagrangian will be particularly discussed.
The present investigation is unlike the work [11], where the onset of chaos in the 2PN Lagrangian formulation
was mainly shown. It is also unlike the paper [33], where the effect of the orbit-spin coupling on the strength of chaos
caused by the spin-spin coupling  was considered but the 1PN and 2PN non-spin terms were not included in the Lagrangian.

In our numerical computations, the velocity of light $c$ and the
constant of gravitation $G$ are taken as geometrized units, $c=G=1$.

\section{Post-Newtonian approaches}

Suppose that compact binaries have masses $M_1$ and $M_2$, and then their
total mass is $M=M_1+M_2$. Other parameters are mass ratio $\beta={M_1}/{M_2}$, reduced mass
$\mu={M_1}{M_2} /{M}$ and mass parameter $\eta=\mu/M=\beta/(1+\beta)^2$. In ADM center-of-mass coordinate
system, $\mathbf{r}$ is a relative
position of body 1 to body 2,
and $\mathbf{v}$ is a velocity of body 1 relative to the center of mass.
Let unit radial vector be
$\mathbf{n}=\mathbf{r}/r$ with radius $r=|\mathbf{r}|$.
The evolution of spinless compact binaries can be described by
the following PN Lagrangian formulation
\begin{equation}
L_{0} =L_N +\frac{1}{c^2}L_{1PN} + \frac{1}{c^4}L_{2PN}.
\end{equation}
The above three parts are the Newtonian term $L_N$, first order PN contribution  $L_{1PN}$
and second order  PN contribution  $L_{2PN}$. They are expressed in Ref. [34] as
\begin{equation}
L_N=\frac{1}{r}+\frac{\mathbf{v}^2}{2},
\end{equation}
\begin{eqnarray}
L_{1PN} &=& \frac{1}{8}(1-3\eta){\mathbf{v}^4}+\frac{1}{2}[(3+\eta){\mathbf{v}^2} +\eta(\mathbf{n}\cdot\mathbf{v})^2]\frac{1}{r} \nonumber \\
& & -\frac{1}{2r^2},
\end{eqnarray}

\begin{eqnarray}
 L_{2PN} &=&
 \frac{1}{16}(1-7\eta+13{\eta}^2){\mathbf{v}^6}+\frac{1}{8}[(7-12\eta-9{\eta}^2){\mathbf{v}^4} \nonumber \\
 & &   +(4-10\eta)\eta(\mathbf{n}\cdot\mathbf{v})^2{\mathbf{v}^2}+3{\eta}^2(\mathbf{n}\cdot\mathbf{v})^4]
  \frac{1}{r} \nonumber
  \\
 &&  +\frac{1}{2}[(4-2\eta+{\eta}^2)\mathbf{v}^2+3\eta(1+\eta)(\mathbf{n}\cdot\mathbf{v})^2]\frac{1}{r^2} \nonumber
  \\
 && +\frac{1}{4}(1+3\eta)\frac{1}{r^3}.
\end{eqnarray}
In fact, these dimensionless equations deal with the use of scale transformation:
$r\rightarrow GMr$, $t\rightarrow GMt$ and $L_0\rightarrow \mu L_0$.
In terms of Legendre transformation $H_0=\mathbf{p}\cdot\mathbf{v}-L_0$ with momenta
$\mathbf{p}=\partial L/\partial\mathbf{v}$,
we have the following Hamiltonian
\begin{equation}
H_{0} =H_N +\frac{1}{c^2}H_{1PN} + \frac{1}{c^4}H_{2PN},
\end{equation}
where these sub-Hamiltonians are
\begin{equation}
H_N = \frac{\mathbf{p}^2}{2}-\frac{1}{r},
\end{equation}
\begin{eqnarray}
H_{1PN} &=& \frac{1}{8}(3\eta-1)\mathbf{p}^4-\frac{1}{2}[(3+\eta)\mathbf{p}^2
+\eta(\mathbf{n}\cdot\mathbf{p})^2]\frac{1}{r} \nonumber \\
&& +\frac{1}{2{r}^2},
\end{eqnarray}
\begin{eqnarray}
H_{2PN}=
& &\frac{1}{16}(1-5\eta+5{\eta}^2)\mathbf{p}^6+\frac{1}{8}[(5-20\eta-3{\eta}^2)\mathbf{p}^4 \nonumber
\\  &&
 -2{\eta}^2(\mathbf{n}\cdot\mathbf{p})^2\mathbf{p}^2-3{\eta}^2(\mathbf{n}\cdot\mathbf{p})^4]\frac{1}{r}
 +\frac{1}{2}[(5+8\eta) \nonumber   \\
 && \mathbf{p}^2+3\eta(\mathbf{n}\cdot\mathbf{p})^2]\frac{1}{r^2}
 -\frac{1}{4}(1+3\eta)
\frac{1}{r^3}.
\end{eqnarray}
The two Hamiltonians $H_{1PN}$ and $H_{2PN}$ are the results of [34]. Besides them, other higher-order PN terms
can be derived from the Lagrangian $L_0$. For example, a third order PN sub-Hamiltonian was given in [26] by
\begin{eqnarray}
H_{3PN}=& &
\frac{3}{16}(-\eta+7{\eta}^2-12{\eta}^3)\mathbf{p}^8+\frac{1}{8r}[(2-7\eta +3{\eta}^2 \nonumber\\
& & +30{\eta}^3)\mathbf{p}^6+(4\eta -11{\eta}^2 +36{\eta}^3)(\mathbf{n}\cdot\mathbf{p})^2 \mathbf{p}^4 \nonumber\\
& & +6({\eta}^2-3{\eta}^3)(\mathbf{n}\cdot\mathbf{p})^4\mathbf{p}^2]
+\frac{1}{4{r}^2}[(5+18\eta \nonumber\\
& & +21{\eta}^2-9{\eta}^3)\mathbf{p}^4 +4(5{\eta}^2+2{\eta}^3)(\mathbf{n}\cdot\mathbf{p})^4 \nonumber\\
& & +(14\eta-7{\eta}^2-27
{\eta}^3)(\mathbf{n}\cdot\mathbf{p})^2\mathbf{p}^2] \nonumber\\
& & +\frac{1}{2{r}^3}[(-3-31\eta-7{\eta}^2+{\eta}^3)
\mathbf{p}^2 \nonumber\\
& & +(-\eta-{\eta}^2+7{\eta}^3) (\mathbf{n}\cdot\mathbf{p})^2].
\end{eqnarray}
It is obtained from coupling of the 1PN term $L_{1PN}$ and the 2PN term $L_{2PN}$.
Since the difference between the 2PN Lagrangian $L_0$ and the 2PN Hamiltonian $H_0$ is at least 3PN order,
the two PN approaches are not exactly equivalent. Clearly, a Hamiltonian that is equivalent to the 2PN
Lagrangian\footnote{Here, the equations of motion derived from this Lagrangian are required
to arrive at a higher enough order or an infinite order.} cannot be at second order but
should be at a higher enough order or an infinite order. This is one of the main results of [26].

When the two bodies spin, some spin effects should be considered. Now, the leading-order spin-spin coupling
interaction $L_{2ss}$ [35], as one kind of spin effect, is included in the Lagrangian $L_0$. In this sense, the obtained
Lagrangian becomes
\begin{eqnarray}
L_1=L_0+L_{2ss},
\end{eqnarray}
where
\begin{eqnarray}
& & L_{2ss} = -\frac{1}{2r^3}[\frac{3}{r^2}({\mathbf{S}_0}\cdot\mathbf{r})^2-{\mathbf{S}_0}^2], \\
& & \mathbf{S} = (2+\frac{3}{2\beta})\mathbf{S}_1+(2+\frac{3\beta}{2})\mathbf{S}_2, \nonumber \\
& & \mathbf{S_0} = (1+\frac{1}{\beta})\mathbf{S}_1+(1+\beta)\mathbf{S}_2. \nonumber
\end{eqnarray}
Note that each spin variable is dimensionless, namely, $\mathbf{S}_i = {S_i}\hat{\mathbf{S}}_i$
with spin magnitude $S_i=\chi_i {M_i}^2/M^2 ~(0 \leq \chi_i \leq 1)$ and 3-dimensional unit vector
$\hat{\mathbf{S}}_i$. Because $L_{2ss}$ is independent of velocity, it does not couple the 1PN term $L_{1PN}$ or the 2PN term $L_{2PN}$
via the Legendre transformation. It is only converted to a spin-spin coupling Hamiltonian
\begin{eqnarray}
H_{2ss} = -L_{2ss}.
\end{eqnarray}
Adding this term to the Hamiltonian $H_0$, we have the following Hamiltonian
\begin{eqnarray}
H_1=H_0+H_{2ss}.
\end{eqnarray}

When another kind of spin effect, the leading-order spin-orbit coupling $L_{1.5so}$ [35], is further
included in the above-mentioned Lagrangian $L_1$, we obtain a Lagrangian formulation as follows:
\begin{eqnarray}
&& L_2 = L_1+L_{1.5so}, \\
& & L_{1.5so} = -\frac{1}{r^3}\mathbf{S}\cdot(\mathbf{r}\times\mathbf{v}).
\end{eqnarray}
Unlike $L_{2ss}$, $L_{1.5so}$ depends on the velocity. Therefore, the Legendre transformation results in
the appearance of the leading-order spin-orbit Hamiltonian $H_{1.5so}$ obtained from the coupling of the Newtonian term  $L_{N}$
and the spin-orbit term $L_{1.5so}$, the next-order spin-orbit Hamiltonian $H_{2.5so}$ obtained from the coupling of the 1PN term  $L_{1PN}$
and the spin-orbit term $L_{1.5so}$ and the next-order spin-spin
Hamiltonian $H_{3ss}$ obtained from the coupling of the leading-order spin-orbit term $L_{1.5so}$
and itself. These Hamiltonians are written in [26] as
\begin{eqnarray}
&& H_{1.5so} = -L_{1.5so}|_{\mathbf{v}\rightarrow\mathbf{p}}, \\
&& H_{2.5so} =
\frac{1}{{r}^3}(\frac{3\eta-1}{2}\mathbf{p}^2-\frac{3+\eta}{r})\mathbf{S}\cdot
(\mathbf{r\times\mathbf{p}}), \\
&& H_{3ss} =
\frac{1}{2{r}^6}[r^{2}\mathbf{S}^2-(\mathbf{S\times\mathbf{r})}^2].
\end{eqnarray}
We take three Hamiltonians:
\begin{eqnarray}
H_2 &=& H_1+H_{1.5so}, \\
H_3 &=& H_2+H_{2.5so} +H_{3PN}, \\
H_4 &=& H_3+H_{3ss}.
\end{eqnarray}
For the three Hamiltonians, $H_4$ is the best approximation to the Lagrangian $L_2$
although the former is not exactly equivalent to the latter. As an important point to note, $L_2$
exhibits chaos when the two objects spin and the spin effects are $L_{1.5so}$ but are not $L_{2ss}$.
This is because many higher-order spin-spin couplings (such as $H_{3ss}$), included in a
higher enough order Hamiltonian equivalent to $L_2$ (with its equations of motion to another
higher enough order), make this equivalent Hamiltonian non-integrable [26]. However, any PN
conservative Lagrangian approach is always integrable and non-chaotic when only one body
spins and the spin effects are not restricted to the spin-orbit couplings. This is due
to integrability of the equivalent Hamiltonian [30, 31].

Clearly, the above-mentioned same PN order Lagrangian and Hamiltonian approaches (e.g. $L_2$ and $H_2$)
are typically nonequivalent, and therefore they both have different dynamics to a large extent.
Now, we are mainly interested in knowing how the spin-orbit coupling $L_{1.5so}$ affects the chaotic behavior yielded by
the spin-spin coupling $L_{2ss}$ in the above Lagrangians. In other words, we should compare
differences between the $L_1$ and $L_2$ dynamics. Considering that the 2PN spin-spin coupling $H_{2ss}$ equivalent to $L_{2ss}$
and the 3PN spin-spin coupling $H_{3ss}$ associated to $L_{1.5so}$ play an important role in
the onset of chaos in the Hamiltonians, we should also focus on differences between the $H_3$ and $H_4$ dynamics.
These discussions await the following numerical simulations.

\section{Numerical comparisons}

Let us take initial conditions $\mathbf{r}(0)=(17.04,0,0)$,
$\mathbf{v}(0)=(0,0.094,0)$, dynamical parameters $\chi_1=\chi_2=1$, $\beta=0.79$, and
initial unit spin vectors
\begin{eqnarray}
&& \mathbf{\hat{S}}_1=(0.1,0.3,0.8)/\sqrt{0.1^{2}+0.3^{2}+0.8^{2}}, \nonumber \\
&& \mathbf{\hat{S}}_2=(0.7,0.3,0.1)/\sqrt{0.7^{2}+0.3^{2}+0.1^{2}}. \nonumber
\end{eqnarray}
An eighth- and ninth-order
Runge-Kutta-Fehlberg  algorithm of variable
step sizes [RKF8(9)] is used to solve the related Lagrangian and Hamiltonian systems.
This algorithm is highly precise. In fact, it gives an order of $10^{-11}$ to the accuracy of the Lagrangian $L_2$
and an order of $10^{-12}$ to the accuracy of the Hamiltonian $H_2$, as shown in Fig. 1.
Here, the errors of $L_2$ and $H_2$ were estimated according to two different integration paths.
The error of $L_2$, i.e., the error of $H_2$, was obtained by  applying the RKF8(9) to solve the 2PN Lagrangian equations of $L_2$,
while the error of $H_2$ was given by applying the RKF8(9) to solve the 2PN Hamiltonian equations of $H_2$.
Although the errors have secular changes, they are so small that the obtained numerical results should be reliable.

The evolution of the orbit in Fig. 2 demonstrates that the same order Lagrangian and Hamiltonian formulations $L_2$ and $H_2$
diverge quickly from the same starting point. This supports again the general result of [26] on the non-equivalence of
the PN Lagrangian and Hamiltonian approaches at the same order.

For the given orbit, we investigate dynamical differences among some  Lagrangian and Hamiltonian formulations using
several methods to find chaos.

\subsection{Chaos indicators}

A power spectral analysis method is based on Fourier transformation and gives the distribution of frequencies to
a certain time series. It can roughly detect chaos from order. Discrete frequencies are usually regarded as power spectra of
regular orbits, whereas continuous frequencies are generally referred to as power spectra of
chaotic orbits. In light of this criterion, distributions of continuous frequency spectra in Figs. 3(a) and (d)-(f)
seem to show the chaoticity of the Lagrangian $L_1$ and the Hamiltonians $H_2$, $H_3$ and $H_4$. On the other hand,
distributions of discrete frequency spectra in Figs. 3(b) and (c)
describe the regularity  of the Hamiltonians $H_1$ and the Lagrangian $L_2$. It is sufficiently proved that
the same PN order Lagrangian and Hamiltonian approaches ($L_1$ and $H_1$, $L_2$ and $H_2$) have different dynamics.
It is worth emphasizing that the method of power spectra is
ambiguous to differentiate among complicated
periodic orbits, quasiperiodic orbits and weakly chaotic
orbits. Therefore, more reliable qualitative methods are necessarily used.

As a common method to distinguish chaos from order, a Lyapunov exponent
characterizes the average exponential deviation of two nearby orbits.
The variational method and the two-particle one are two
algorithms for calculating the Lyapunov exponent [36, 37]. Although the latter method is less rigorous than
the former method, its application to a complicated system is more convenient.
For the use of the two-particle method, the principal Lyapunov exponent is
defined as
\begin{equation}
\lambda=\lim_{t\rightarrow\infty} \frac{1}{t}\ln\frac{d(t)}
{d(0)},
\end{equation}
where $d(0)$ and $d(t)$
are separations between two neighboring orbits at times 0 and $t$, respectively.
The best choice of the initial separation  $d(0)$ is an order of $10^{-8}$ under the circumstance of double precision [36].
In addition, avoiding saturation of orbits needs renormalization.
A global stable system\footnote{As an example, a global stable binary system in Ref. [30] means that
the two objects do not run to infinity, and do not merge, either. In fact, the binaries move in a bounded region.} is chaotic if $\lambda$
reaches a stabilizing positive value, whereas it is ordered
when $\lambda$ tends to zero. In terms of this, it can be seen clearly from Fig. 4 that the four approaches $L_1$, $H_2$, $H_3$ and $H_4$
with positive Lyapunov exponents are chaotic, and the two formulations $H_1$ and $L_2$ with zero Lyapunov exponents are regular.
Note that each of these values of Lyapunov exponents is given after integration time
$2\times 10^5$. Obtaining a reliable stabilizing limit value of Lyapunov exponent usually costs an extremely
expensive computation [20].

A quicker method to find chaos is a fast Lyapunov indicator (FLI) [38, 39].
It was originally calculated using the length of a tangential vector and renormalization is unnecessary.
Similar to the Lyapunov exponent, this indicator can be further developed with the separation between two nearby trajectories [40].
The modified indicator is of the form
\begin{equation}
\text{FLI}=\log_{10}\frac{d(t)}
{d(0)}.
\end{equation}
An appropriate choice for renormalization within a reasonable amount of time span
is important. See Ref. [40] for more details of this indicator.
The FLI increases exponentially with time for a chaotic orbit, but algebraically for a regular
orbit. The completely different time rates are used to distinguish between the two
cases. Based on this point, the dynamical behaviors of the six approaches $L_1$, $L_2$, $H_1$, $H_2$, $H_3$ and $H_4$
can be described by the FLIs in Fig. 5. These results are the same as those given by the Lyapunov exponents in Fig. 4.
Here each FLI was obtained after $t=5\times 10^4$.
In this sense, the FLI is indeed a faster method to identify chaos than the Lyapunov exponent.
Because of this advantage, the method of FLIs is widely used to sketch out the global structure of phase space or
to provide some insight into dependence of chaos on single physical parameter or initial condition [23].

\begin{table}
\caption{Chaotic parameter spaces of each approach}
\label{Tab1}
\begin{tabular*}{\columnwidth}{@{\extracolsep{\fill}}ll@{}}
\hline
Approach & Mass Ratio $\beta$  \\
\hline
$L_{1}$       &  0.05, 0.06, 0.08, [0.11,0.16], [0.18,0.83], \\
                         &  0.86, 0.86, [0.89,0.92], 0.94               \\ \hline
$H_{1}$       &  [0.12,0.19], [0.23,0.27], 0.30, \\
                         &  [0.33,0.34], [0.40,0.42], [0.62,0.64]              \\ \hline
$L_{2}$       &  0.03, 0.05, [0.09,0.41], [0.43,0.45], 0.47, 0.54    \\ \hline
$H_{2}$       & 0.09, [0.17,0.22], [0.24,0.99]               \\ \hline
$H_{3}$ & [0.09,0.22], [0.24,0.65], [0.67,0.97]\\ \hline
$H_{4}$    &[0.09,0.10], [0.15,0.17], 0.19, 0.48, 0.62, 0.65, \\
                               & 0.68, 0.75, 0.79, 0.82, [0.84,0.91], 0.93      \\ \hline
\end{tabular*}
\end{table}

\subsection{Effects of varying the mass ratio on chaos}

Fixing the above initial conditions, initial spin configurations and spin parameters ($\chi_1$ and $\chi_2$),
we let the mass ratio $\beta$ run from 0 to 1 in increments of 0.01.
For each value of $\beta$, FLI is obtained after integration time $t=5\times10^4$.
In this way, we plot Fig. 6 in which dependence of FLI for each PN approach on $\beta$ is described.
It is found through a number of numerical tests that
7.5 is a threshold of FLI to distinguish between regular and chaotic orbits.
A global stable orbit is chaotic if its FLI is larger than the threshold,
but ordered if its FLI is smaller than the threshold. In this sense, this figure
shows clearly the correspondence between the mass ratio and the orbital dynamics.

It is shown sufficiently that the PN Lagrangian and Hamiltonian approximations at the same order,
($L_1$, $H_1$) and ($L_2$, $H_2$), have different dynamical behaviors in significant measure.
More details on the related differences are listed in Table 1. It can also be seen that the chaotic parameter space of
$L_1$ is larger than that of $L_2$.
That means that the spin-orbit coupling $L_{1.5so}$
makes many chaotic orbits in the system $L_1$ evolve into
ordered orbits.\footnote{The chaoticity of the system $L_1$ is due to the spin-spin coupling $L_{2ss}$.
$L_2$ minus $L_1$ is $L_{1.5so}$.} In other words, the probability of the occurrence of chaos in the system $L_1$ without $L_{1.5so}$
is large, but that in the system $L_2$ with $L_{1.5so}$
is small. Thus, $L_{1.5so}$ plays an important role in weakening or suppressing the chaotic behaviors yielded by  $L_{2ss}$.
Here, the chaotic behaviors weakened (or suppressed) do not mean that an individual orbit must become
from strongly chaotic to weakly chaotic (or from chaotic to nonchaotic) when $L_{1.5so}$ is included in $L_1$.
This orbit may become more strongly chaotic, or may vary from order to chaos. For example, $\beta=0.03$ in Table 1 corresponds to the
regularity of $L_1$ but the chaoticity of $L_2$. In fact, the chaotic behaviors weakened or suppressed mean decreasing
the chaotic parameter space. The result obtained in the general case with the PN terms  $L_{1PN}$ and $L_{2PN}$
is an extension to the special case without the PN terms  $L_{1PN}$ and $L_{2PN}$ in Ref. [33]. As the authors of [33] claimed,
this result is due to different signs of $H_{2ss}$ (equivalent to $L_{2ss}$)  and $H_{3ss}$ (associated to $L_{1.5so}$).
The two spin-spin terms are responsible for causing chaos in the Hamiltonian $H_4$, and $H_{2ss}$
has a more primary contribution to chaos. It is further shown in Fig. 6 and Table 1 that $H_4$ with the inclusion of $H_{3ss}$
has weak chaos and a small chaotic parameter space, compared with $H_3$ with the absence of $H_{3ss}$.
This is helpful for us to explain why $L_{1.5so}$ can somewhat weaken or suppress the chaoticity caused by $L_{2ss}$ in the PN Lagrangian system $L_2$.

\section{Conclusions}

When a Lagrangian function at a certain PN order is transformed into a Hamiltonian function,
many additional higher-order PN terms usually occur. In this sense, the PN Lagrangian and Hamiltonian approaches
at the same order are generally nonequivalent. The equivalence between the Lagrangian formulation and a Hamiltonian system often
requires that the Euler-Lagrangian equations and the Hamiltonian should be up to higher enough orders or an infinite order.

For the Lagrangian formulation of  spinning compact binaries, which includes the Newtonian term, 1PN and 2PN non-spin terms
and 2PN spin-spin coupling, the Legendre transformation gives not only the same order PN Hamiltonian but also many
additional higher-order PN terms, such as the 3PN non-spin term. Therefore, the same order Lagrangian and Hamiltonian approaches
have some different dynamics. This result is confirmed by numerical simulations. This Lagrangian is non-integrable
and can be chaotic under an appropriate circumstance due to the absence of a fifth integral of motion in the equivalent Hamiltonian.
When the 1.5PN spin-orbit coupling is added to the Lagrangian, the 3PN spin-spin coupling appear in the derived Hamiltonian.
The 3PN spin-spin Hamiltonian is small and has different signs compared with the 2PN spin-spin Hamiltonian. In this sense,
the probability of the occurrence of chaos in the Lagrangian formulation without the spin-orbit coupling
is large, whereas that in the Lagrangian formulation with the spin-orbit coupling
is small. That means that
the leading-order spin-orbit coupling can somewhat weaken or suppress the chaos yielded by the leading-order spin-spin coupling
in the PN Lagrangian formulation. Numerical results also support this fact.

\section*{Acknowledgments}
This research has been supported by the Natural Science Foundation of Jiangxi Province under Grant No. [2015] 75, the National Natural
Science Foundation of China under Grant No. 11533004
and the Postgraduate Innovation Special Fund Project of Jiangxi Province
under Grant No. YC2015-S017.

\begin{figure*}[ptb]
\center{
\includegraphics[scale=0.75]{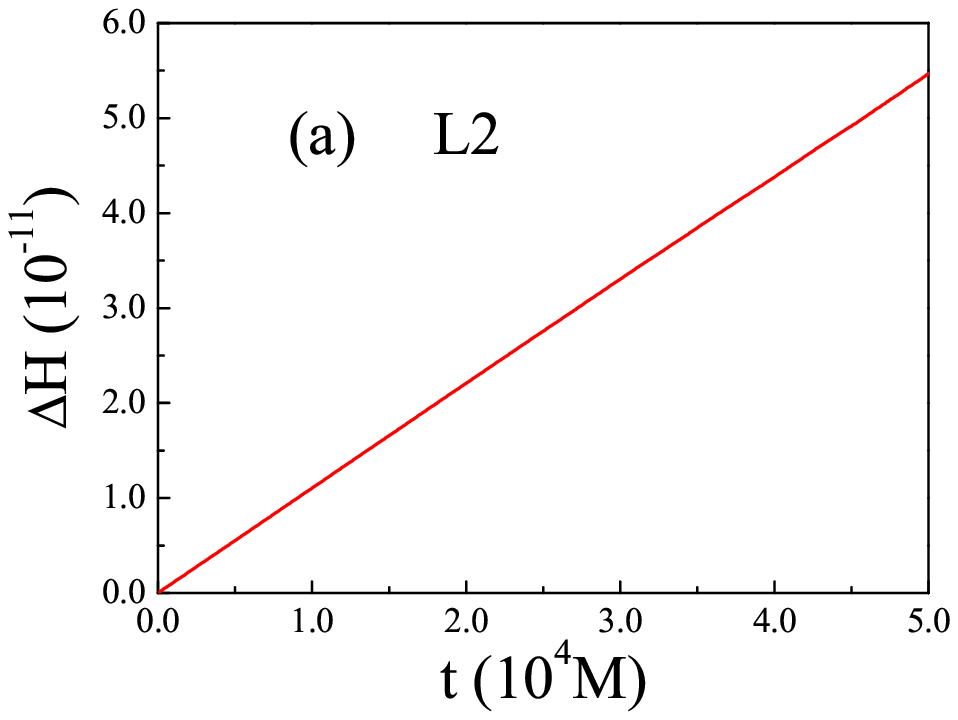}
\includegraphics[scale=0.75]{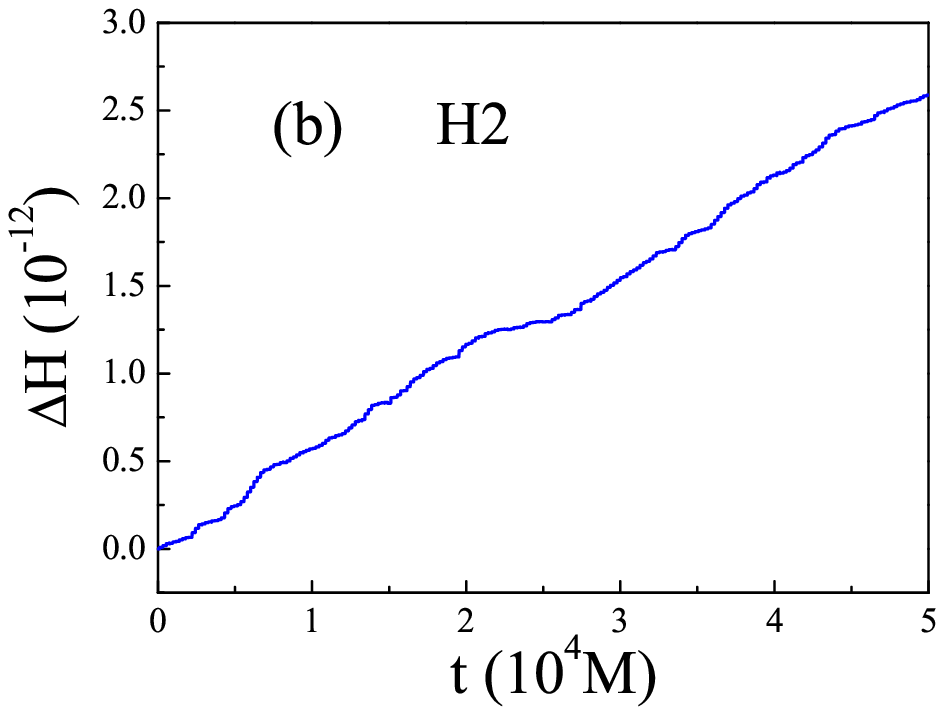}
\caption{(color online) Hamiltonian errors of the 2PN Lagrangian and Hamiltonian approaches, $L_2$ and $H_2$. (a) The RKF8(9) is used
to solve the Euler-Lagrangian equations of $L_2$ so as to obtain the difference $\Delta H$ between the Hamiltonian $H_2(t)$ at time $t$
and the initial Hamiltonian $H_2(0)$. (b) The integrator directly solves the Hamiltonian $H_2$ and obtain the difference $\Delta H$.}
 \label{Fig1}}
\end{figure*}

\begin{figure*}[ptb]
\center{
\includegraphics[scale=0.68]{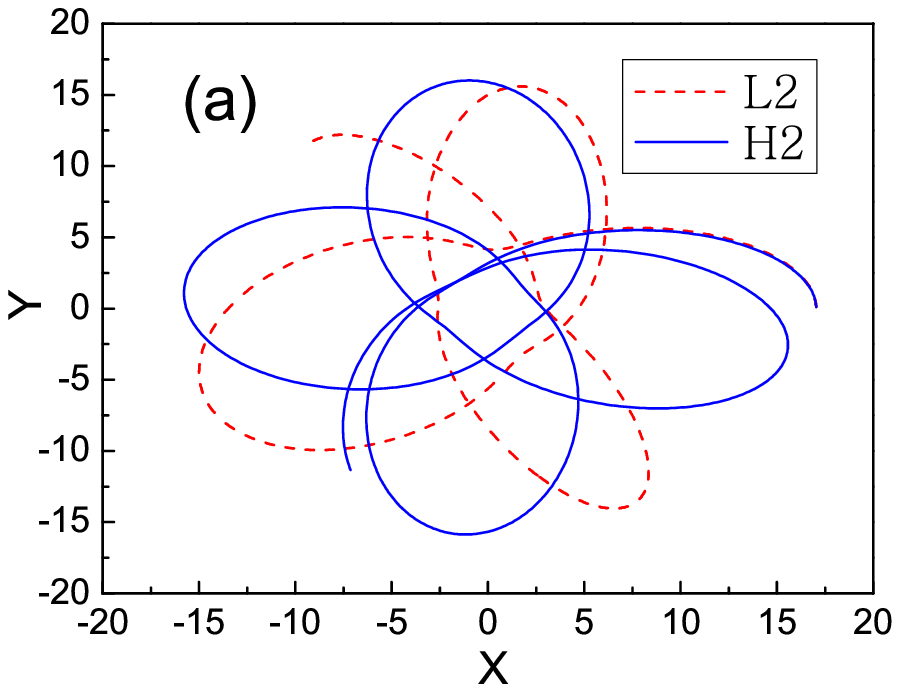}
\includegraphics[scale=0.75]{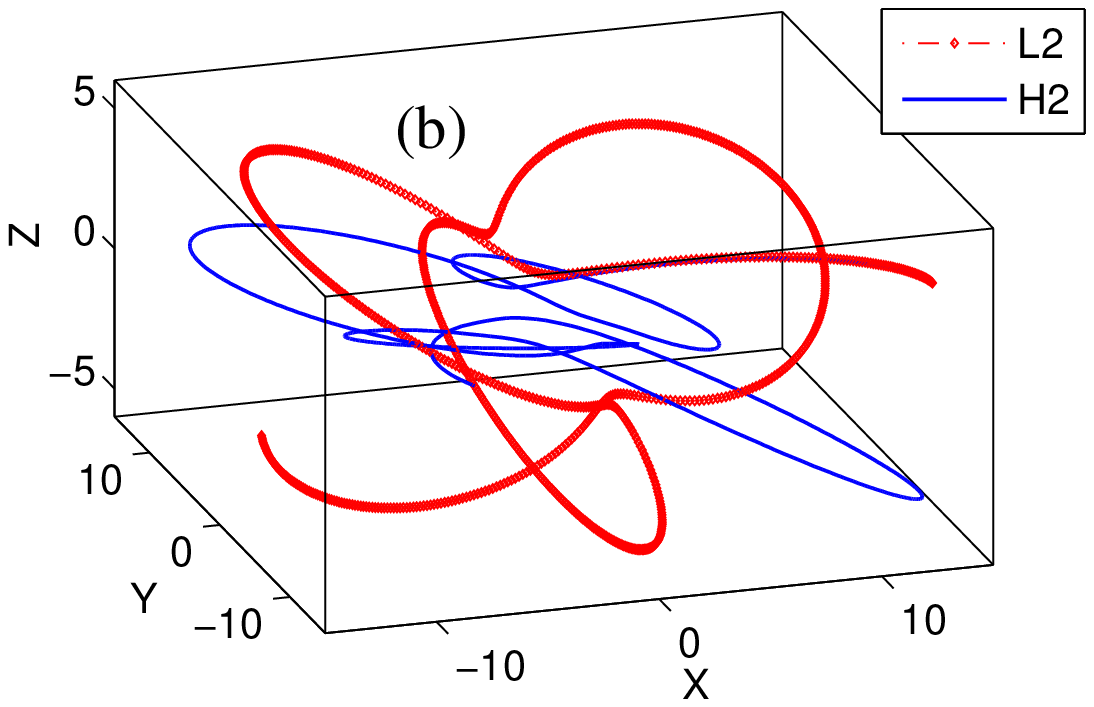}
\caption{(color online) Comparison of orbits in the 2PN
Lagrangian and Hamiltonian approaches, $L_2$ and $H_2$.
(a) The orbits projected onto the $X$-$Y$ plane, and
(b) the orbits in the three-dimensional Euclidean space.}
\label{Fig2}}
\end{figure*}

\begin{figure*}[ptb]
\center{
\includegraphics[scale=0.5]{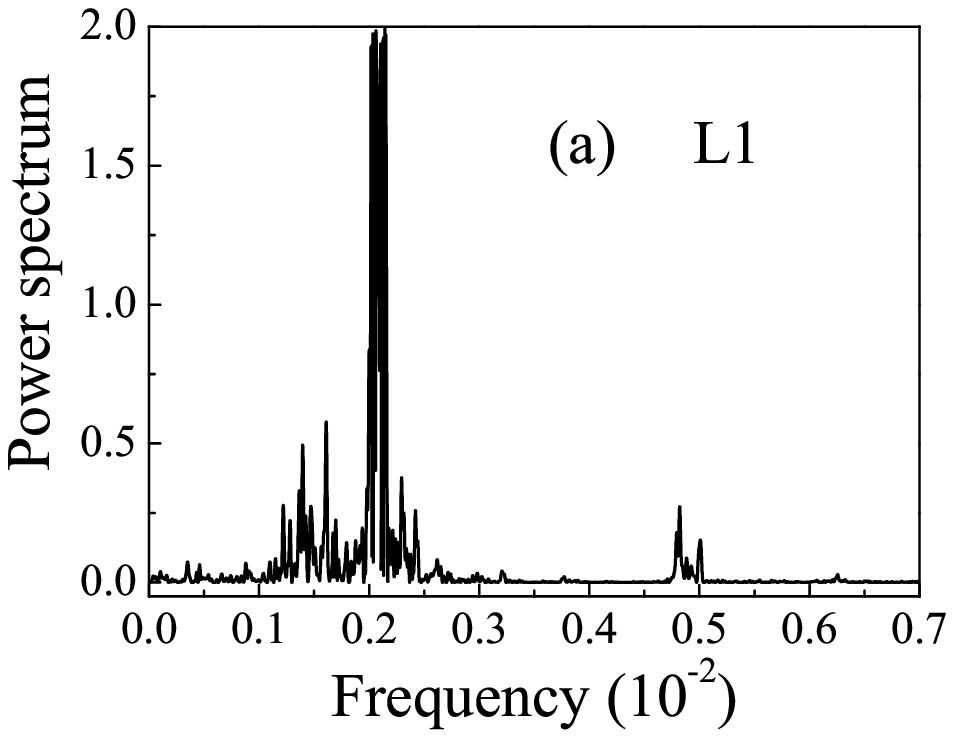}
\includegraphics[scale=0.5]{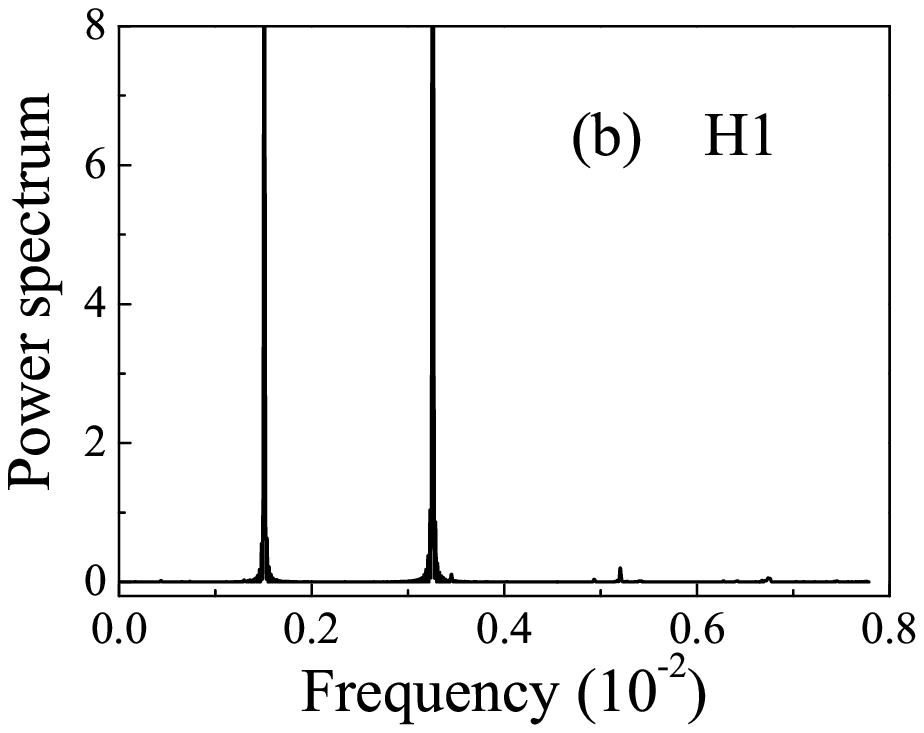}
\includegraphics[scale=0.5]{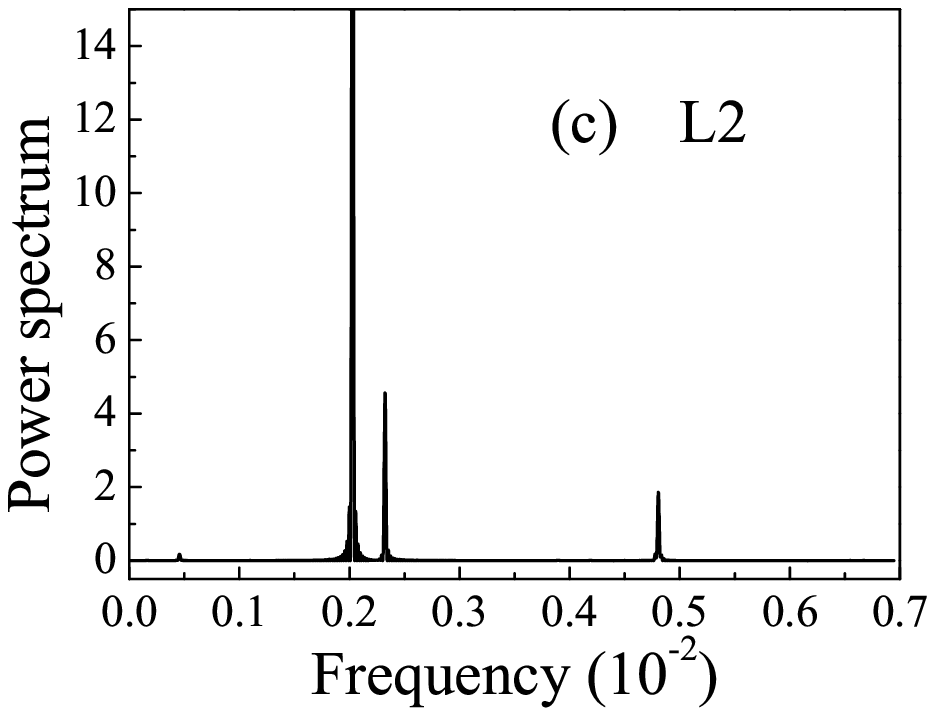}
\includegraphics[scale=0.5]{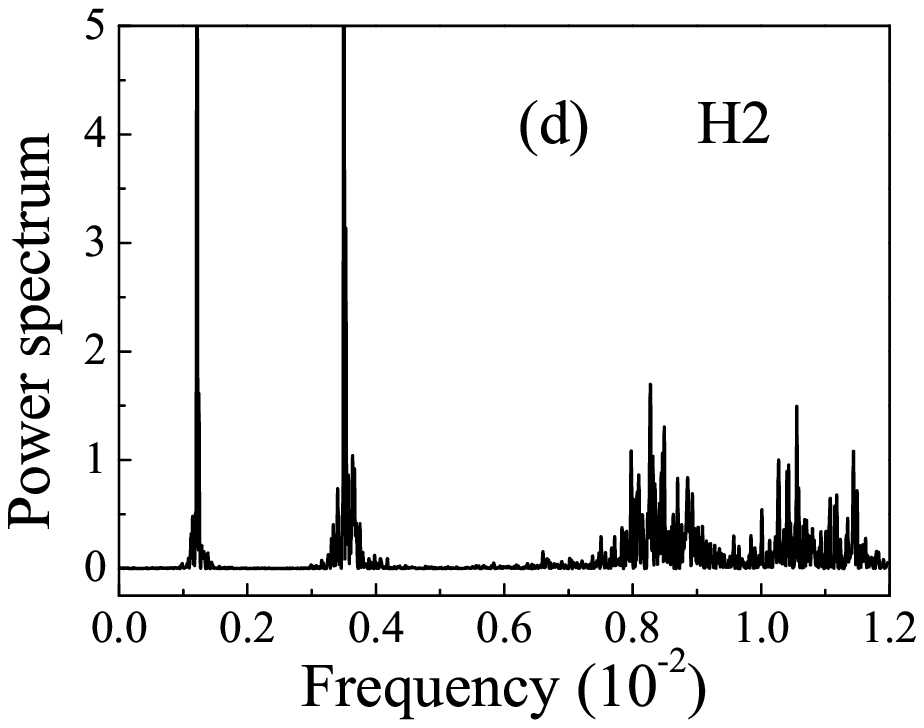}
\includegraphics[scale=0.5]{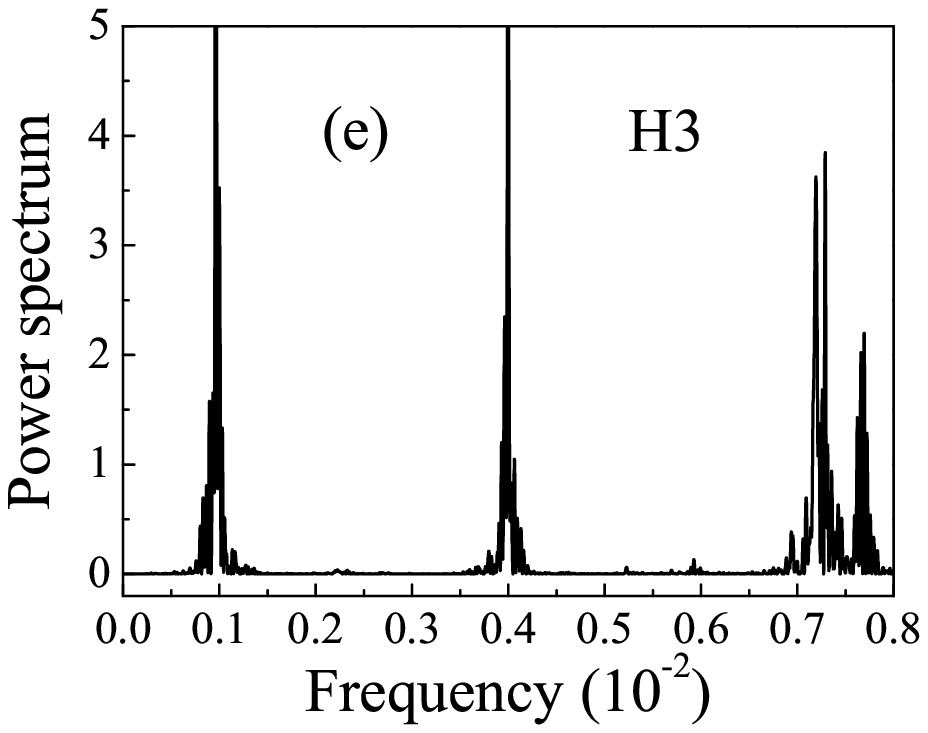}
\includegraphics[scale=0.5]{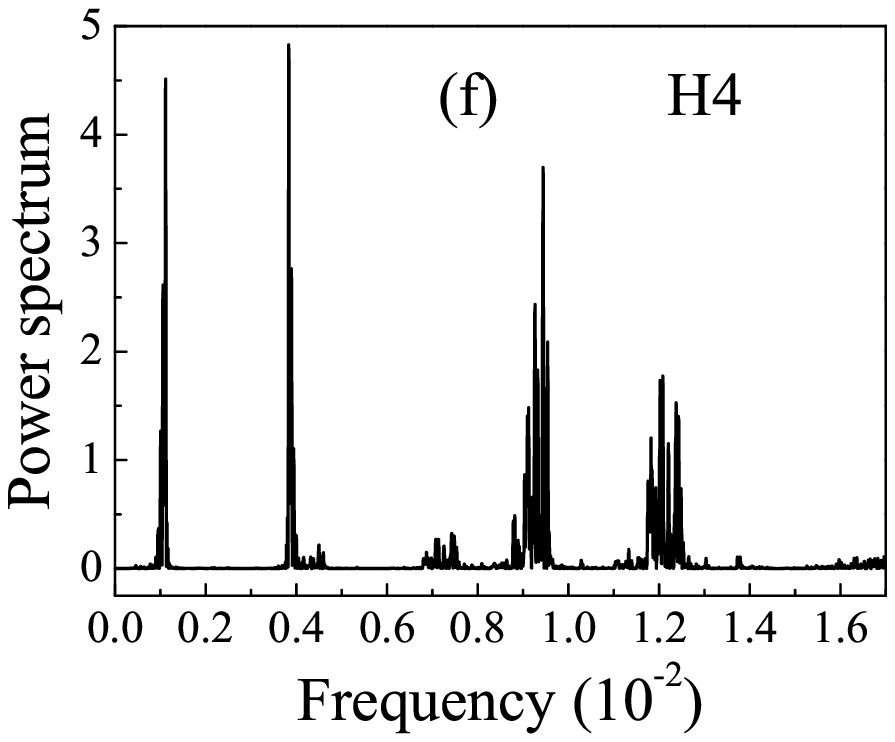}
\caption{Power spectra of six PN approaches.}
\label{Fig3}}
\end{figure*}

\begin{figure*}[ptb]
\center{
\includegraphics[scale=0.7]{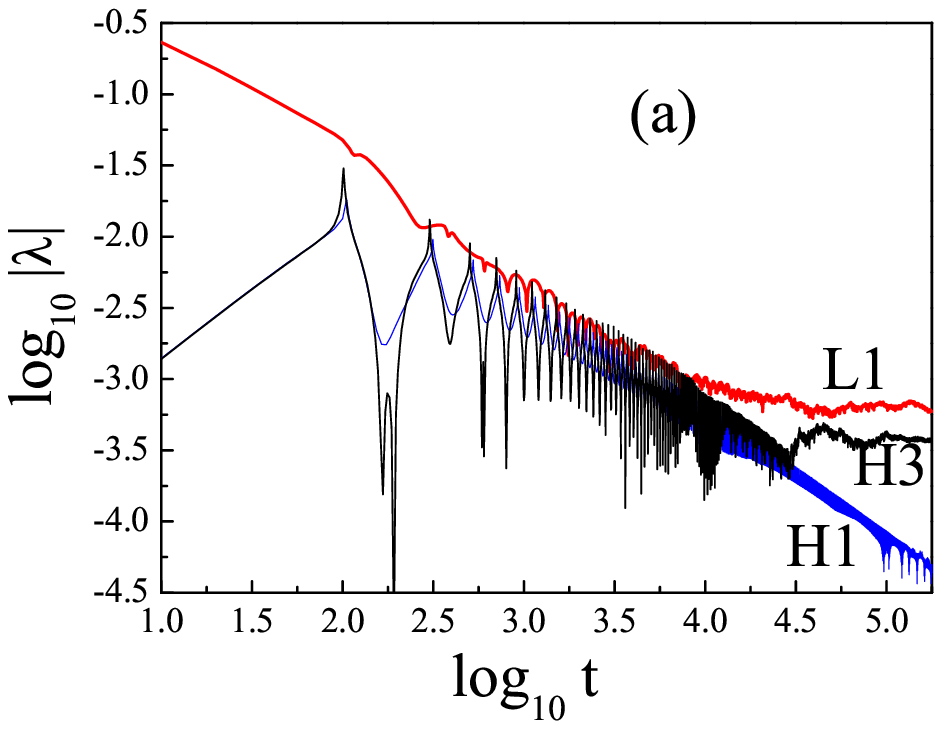}
\includegraphics[scale=0.7]{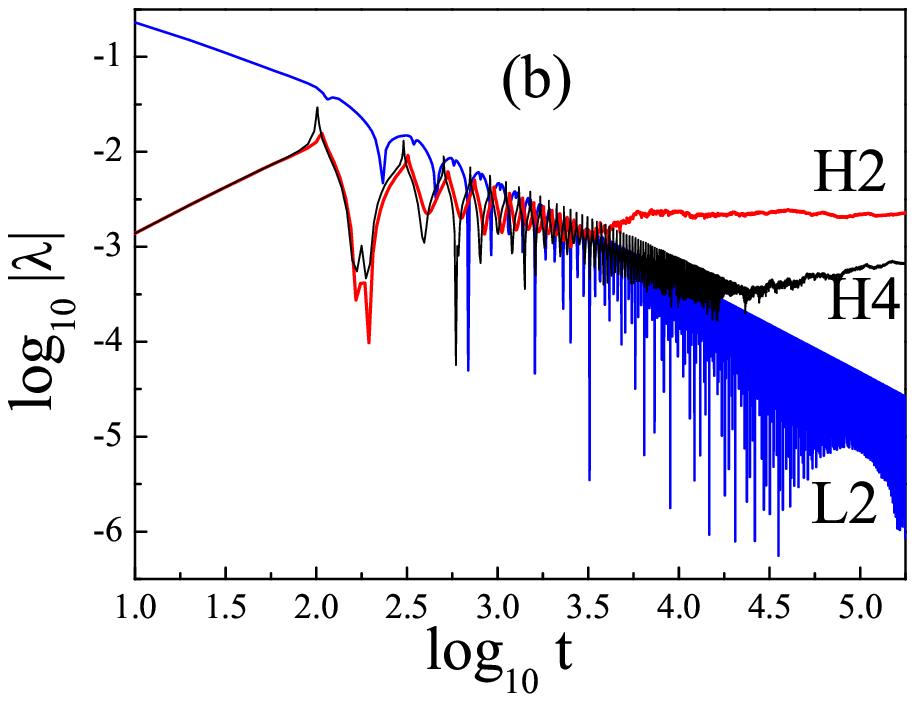}
\caption{(color online) Lyapunov exponents $\lambda$ of six PN approaches.}
\label{Fig4}}
\end{figure*}

\begin{figure*}[ptb]
\center{
\includegraphics[scale=0.7]{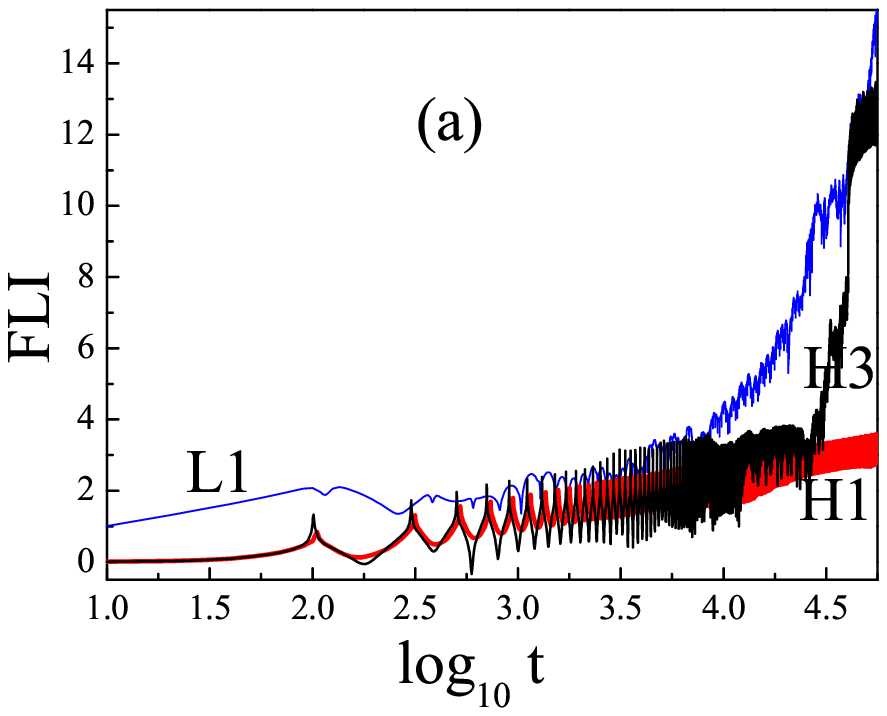}
\includegraphics[scale=0.7]{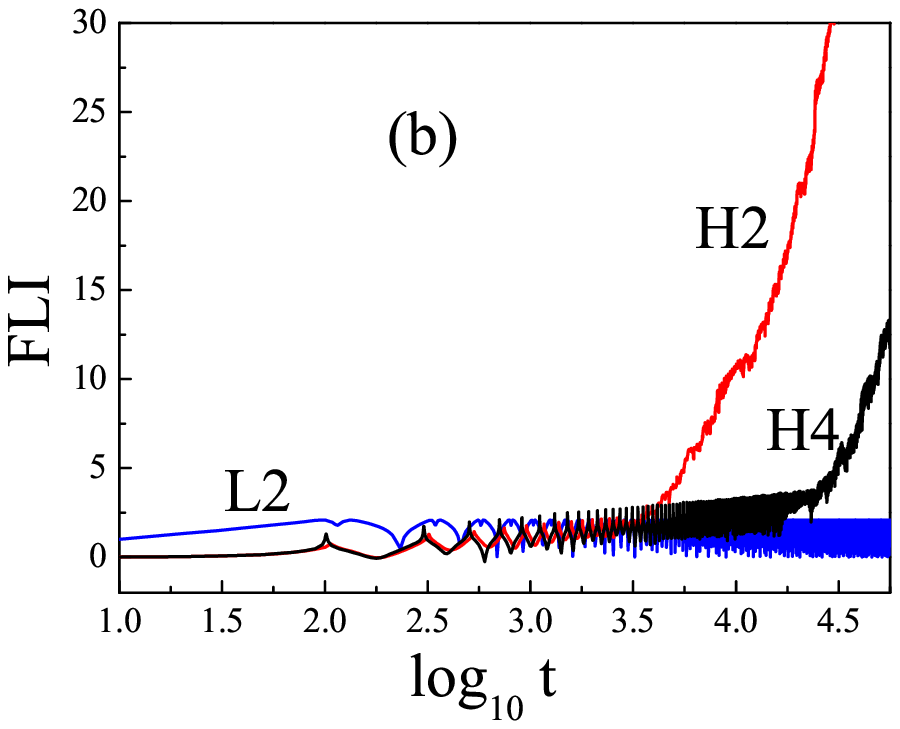}
\caption{(color online) Fast Lyapunov indicators (FLIs) of six PN approaches.}
\label{Fig5}}
\end{figure*}

\begin{figure*}[ptb]
\center{
\includegraphics[scale=0.5]{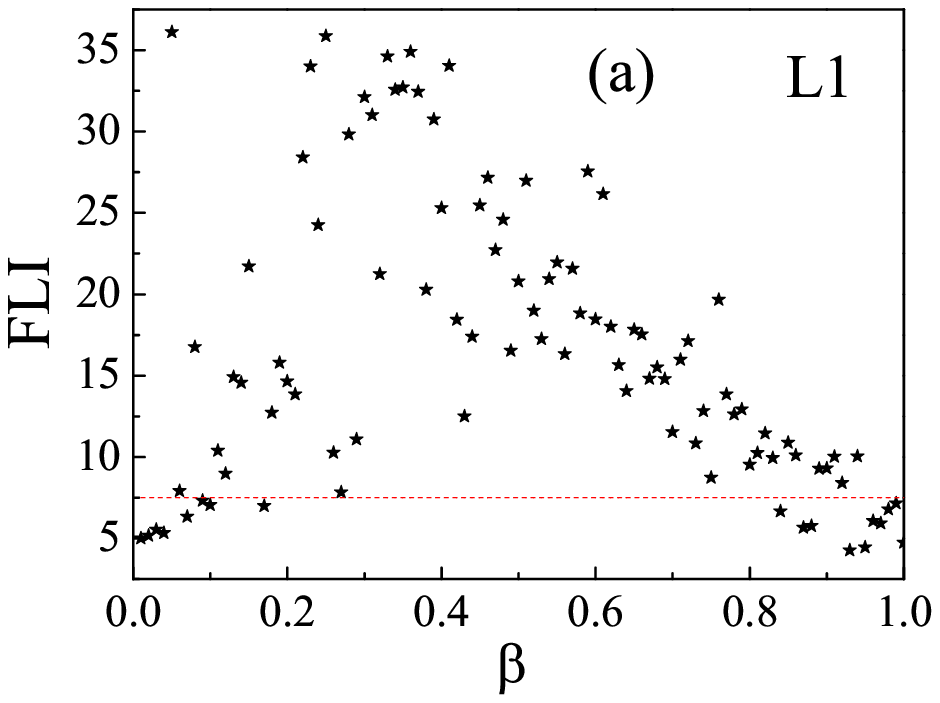}
\includegraphics[scale=0.5]{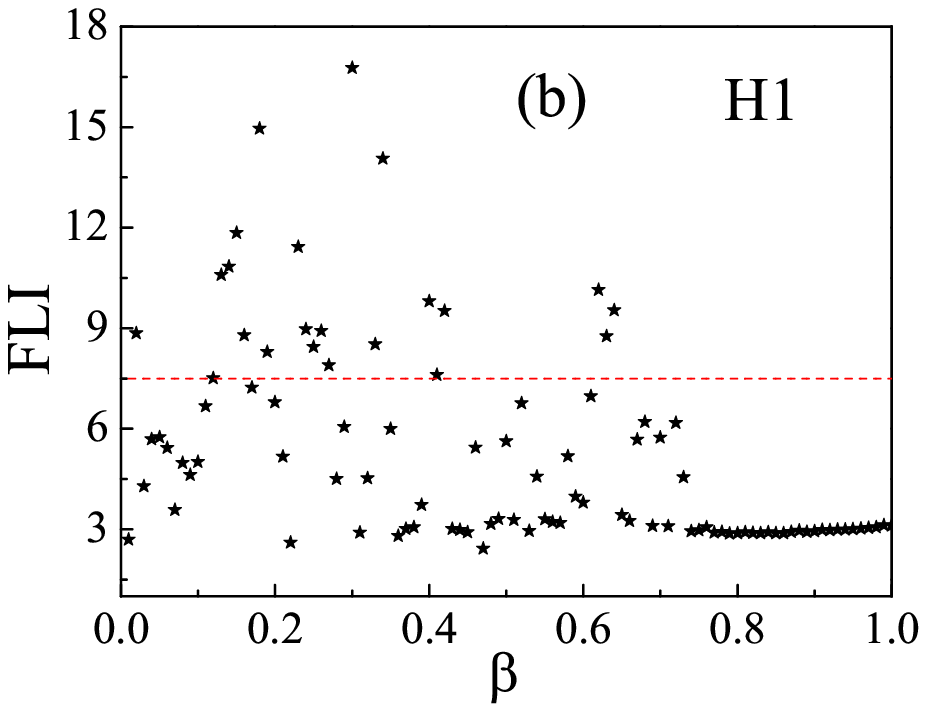}
\includegraphics[scale=0.5]{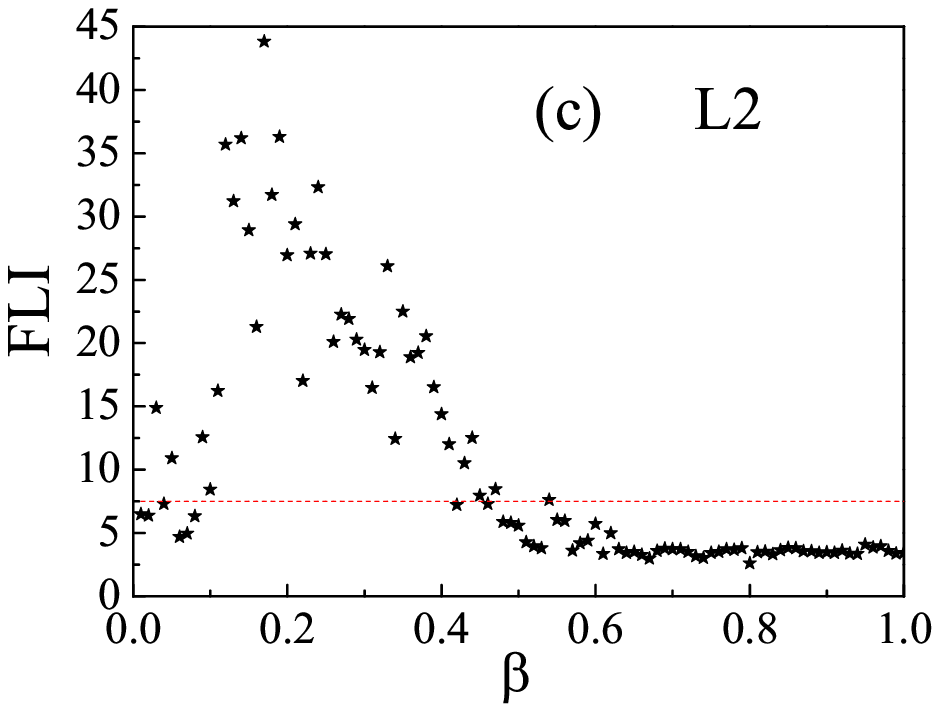}
\includegraphics[scale=0.5]{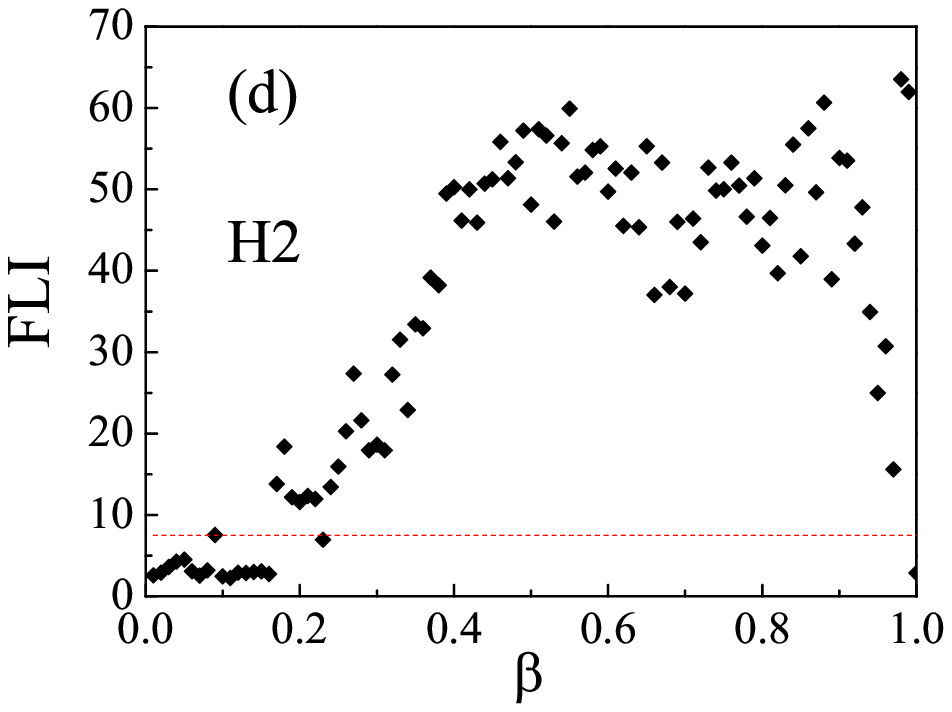}
\includegraphics[scale=0.5]{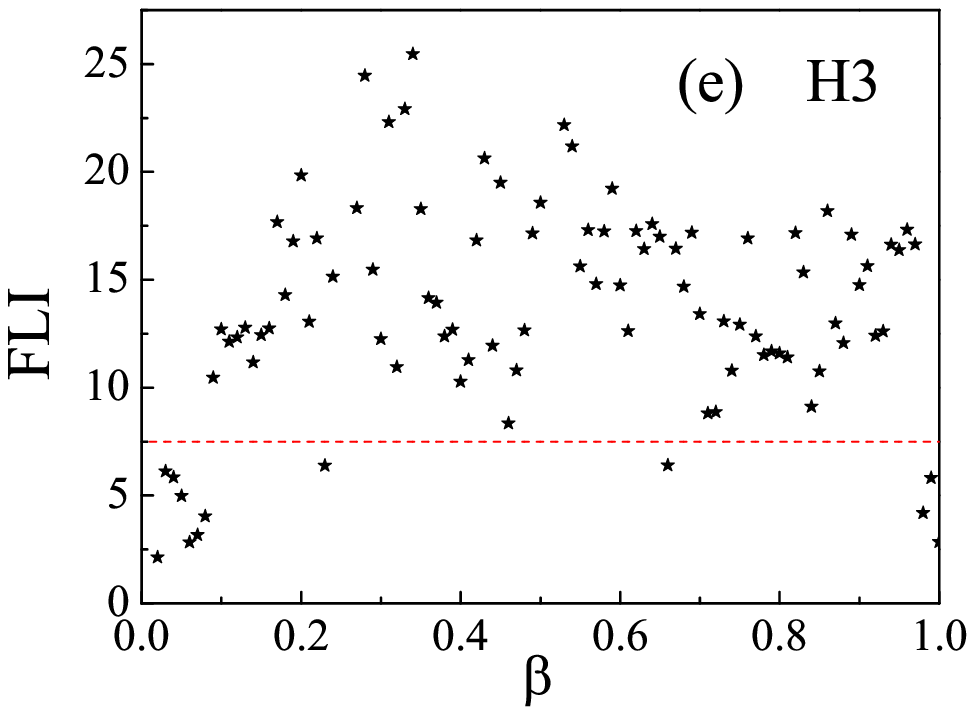}
\includegraphics[scale=0.5]{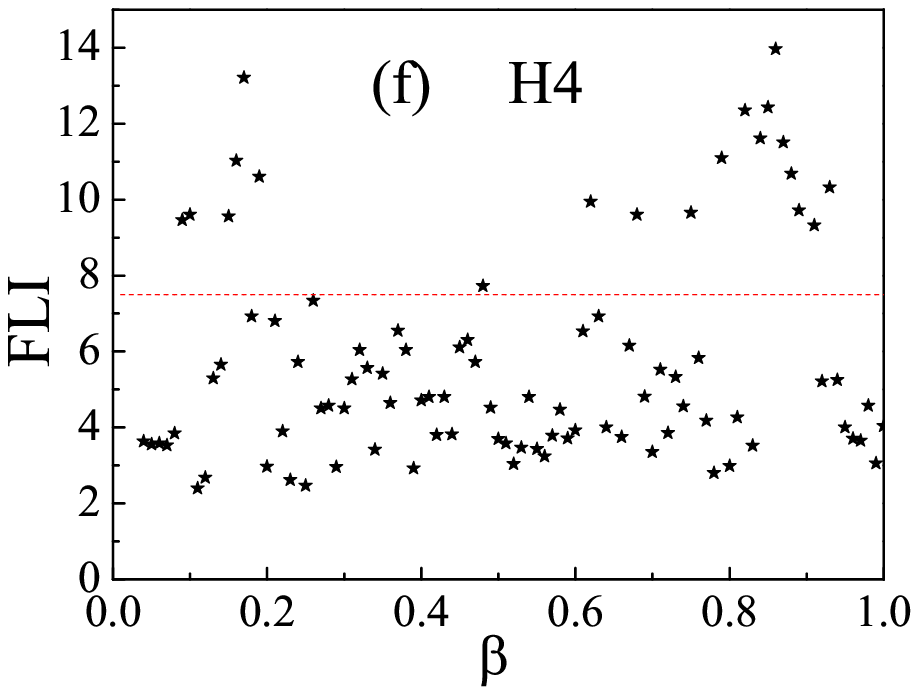}
\caption{(color online) Dependence of FLI on the mass ratio $\beta$ for each approach.}
\label{Fig6}}
\end{figure*}

\end{document}